\documentclass[11pt]{iopart}

\expandafter\let\csname equation*\endcsname\relax
\expandafter\let\csname endequation*\endcsname\relax

\usepackage{amsfonts}
\usepackage{amstext}
\usepackage{amssymb}
\usepackage{bm}
\usepackage{calrsfs}
\usepackage{cite}
\usepackage{dcolumn}
\usepackage{framed}
\usepackage{graphicx}
\usepackage{graphics}
\usepackage[latin1,utf8]{inputenc}
\usepackage{latexsym}
\usepackage{mathtools}
\usepackage{rotating}
\usepackage{url}
\usepackage[dvipsnames]{xcolor}
\usepackage{hyperref}
\hypersetup{colorlinks=true,
            linkcolor=Plum,
            citecolor=Plum,
            filecolor=Plum,
            urlcolor=Plum}
\usepackage{xspace} % Sensible space treatment at end of simple macros
\usepackage{booktabs}

\definecolor {darkgreen}{rgb}{0.2,0.7,0.2}

\newcommand{\gcmc}{\,g/cm$^{3}$}                % per cm-squared
\newcommand{\msun}{\,${\rm M}_  {\odot}\,$}       % solar mass
\newcommand{\ilq}{I-Love-Q\,  }               % I-Love-Q
\newcommand{\veps}{\varepsilon}                 % energy density / c^2
\newcommand{\dd}{{\rm d}}

\newcommand\apj{Astrophys. J.}
\newcommand\nat{Nature}
\newcommand\prd{Phys. Rev. D}

\newcommand\physrep{Phys. Rep.}
\newcommand\aap{A\&A}
\newcommand\aaps{A\&AS}
\newcommand\procspie{Proc. SPIE}

\begin{document}
\title[I-Love-Q to the extreme]{I-Love-Q to the extreme}

\author{%
Hector~O.~Silva$^{1}$
and
Nicol\'as~Yunes$^{1}$
}

\address{$^{1}$~eXtreme Gravity Institute, Department of Physics,
Montana State University, Bozeman, MT 59717, USA.}

\date{\today}

%%%%%%%%%%%%%%%%%%%%%%%%%%%%%%%%%%%%%%%%%%%%%%%%%
\begin{abstract}

% Intro
Certain bulk properties of neutron stars, in particular their moment of inertia, rotational quadrupole moment and tidal Love number, when properly normalized, are related to one another in a nearly equation of state independent way.
% Goals of the paper
The goal of this paper is to test these relations with extreme equations of state at supranuclear densities constrained to satisfy only a handful of generic, physically sensible conditions.
% Class of EoSs
By requiring that the equation of state be (i) barotropic and (ii) its associated speed of sound be real, we construct a piecewise function that matches a tabulated equation of state at low densities, while matching a stiff equation of state parametrized by its sound speed in the high-density region.
% Results of the paper
We show that the I-Love-Q relations hold to 1 percent with this class of equations of state, even in the extreme case where the speed of sound becomes superluminal and independently of the transition density.
% Why I-Love-Q support
We also find further support for the interpretation of the I-Love-Q relations as an emergent symmetry due to the nearly constant eccentricity of isodensity contours inside the star.
% Why this matters.
These results reinforce the robustness of the I-Love-Q relations against our current incomplete picture of physics at supranuclear densities, while strengthening our confidence in the applicability of these relations in neutron star astrophysics.

\end{abstract}

% \pacs{04.50.Kd, 04.70.-s, 04.80.Cc, 04.25.dg}
% \submitto{\CQG}
% \noindent{\it Keywords\/}: general relativity, modified theories of gravity, black holes, continuum spectrum, accretion disks

\maketitle

%%%%%%%%%%%%%%%%%%%%%%%%%%%%%%%%%%%
\section{Introduction}
\label{sec:intro}

Neutron stars are ideal laboratories for probing fundamental physics.
The energy density inside their inner core can be a few times above
nuclear saturation density, where matter transmutes into forms that cannot
be recreated in the laboratory. Their compactness creates large gravitational
fields, which demand the use of relativistic gravity for their modeling.
Their rotation frequency can rival that of the blades of the fastest
professional kitchen blender, while supporting the most extreme magnetic
fields. These are just a few of the many striking features that make neutron
stars invaluable tools to study extreme physics.

An outstanding problems in neutron star astrophysics is the determination of
the equation of state (EoS) of cold-nuclear matter inside the core, which
is a crucial ingredient in the prediction of observables, such as the
(gravitational) mass $M$ and radius
$R$~\cite{Lattimer:2012nd,Miller:2016pom,Watts:2016uzu}.
While the EoS in the crust region is fairly well-understood, there is a
large degree of uncertainty at 1--2 times the nuclear saturation energy
density $\veps_{\rm n} \simeq 150$ MeV fm$^{-3}$ $\simeq 2.67 \times
10^{14}$\gcmc, and this uncertainty increases further at densities
$\simeq 10^{15}$\gcmc~common in the inner core. One could imagine using
precise measurements of the mass of neutron stars to observationally
determine the EoS. Unfortunately, competing EoS models predict neutron
stars that fill a large portion of the mass-radius ($M$--$R$) plane,
introducing degeneracies when converting from observed masses
to constraints on the EoS.

But not all is lost on the observational front. Accurate measurements
of massive $\simeq 2$\msun pulsars~\cite{Demorest:2010bx,Antoniadis:2013pzd}
set a solid lower mass bound that viable EoSs must respect, thus ruling
out a number of candidate EoSs. Similarly, one can use the observed
population of pulsars to place a statistical upper limit on the maximum
mass of neutron stars~\cite{Antoniadis:2016hxz,Alsing2017}, thus
further restricting the space of viable EoSs. Unfortunately, while the
mass of neutron stars can sometimes be accurately measured
(see~\cite{Alsing2017} for an overview), measuring the radius is currently
much more difficult~\cite{Ozel:2016oaf}. Future simultaneous
mass-radius measurements will hopefully tighten the $M$--$R$ relation,
thus placing even stronger constraints on the EoS.

The uncertainties of the EoS also impact our ability to determine
other global properties of neutron stars, which are important in
various astrophysical scenarios. The moment of inertia
determines the spin-orbit coupling correction to the rate of advance of
the periastron in binary pulsars~\cite{Barker1975,Damour1988}, an effect that
may be measurable in the future with the double-pulsar system
PSR J0737-3039~\cite{Burgay:2003jj,Lyne:2004cj,Lattimer:2004nj}.
The stellar ellipticity, the moment of inertia and the quadrupole moment
affect the modeling of x-rays emitted by hot spots on the surface of neutron
stars, as light is affected by the curvature of the exterior spacetime when it
escapes~\cite{Cadeau:2004gm,Cadeau:2006dc,Morsink:2007tv,Psaltis:2013fha,Psaltis:2013zja}.
The quadrupole moment and the tidal Love number (associated
with the tidal deformability of the star) are also important in the modeling of gravitational waves
emitted in the inspiral of binary neutron stars, as these introduce finite-size corrections to models
constructed in the point-particle limit~\cite{Mora:2003wt,Berti:2007cd,Flanagan:2007ix,Read:2009yp,Read:2013zra,Lackey:2013axa,Yagi:2015pkc,Dietrich:2015pxa,DelPozzo:2013ala}.

Uncertainties in the EoS can therefore introduce degeneracies that
limit the amount of information that can be extracted from observations,
unless one finds approximately EoS-independent relations between model
parameters.
Imagine for example that the $f$, $w$ and $p$-mode frequencies
$\omega_{i}$ of pulsating neutron stars are extracted from
future gravitational wave observations. Approximately EoS-independent
relations between these frequencies and the compactness $C \equiv M/R$
or ``average density'' $M/R^3$, could then be used to extract the mass
and the radius of neutron stars from the observation of any two
frequencies~\cite{Andersson:1996pn,Andersson:1997rn}.
In practice, of course, the accuracy of this extraction is limited
not only by the statistical uncertainties in the measurement of the
frequencies, but also potential systematic uncertainties in the degree
of EoS-independence of the relations themselves.

In this spirit,~\cite{Yagi:2013bca,Yagi:2013awa} found that
the moment of inertia $I$, the quadrupole moment $Q$ and
tidal Love number $\lambda$, when properly normalized,
are linked together in a remarkable (nearly) EoS-independent way
(see~\cite{Yagi:2016bkt,Doneva:2017jop} for recent reviews).
The analytic fits obtained
by~\cite{Yagi:2013bca,Yagi:2013awa}, connecting two of the
quantities in the \ilq trio, hold to better than $1\%$
for all EoS considered, covering a very broad spectrum of nuclear
physics models.
These universal relations can be used in a number
of physical scenarios. For instance, take the case of the
PSR J0737-3039 system. If a measurement of the moment of inertia
of Pulsar A is made, knowledge of its mass $M$ and spin, allows us to
automatically infer its quadrupole moment $Q$. The \ilq relations can also
be used to break degeneracies between various parameters in models of
x-ray pulse profiles and gravitational waves. In this way,
these relations reduce the size of the parameter space of the
models and allow the remaining parameters to be more accurately
determined~\cite{Yagi:2016bkt}.

Given the tremendous potential of the \ilq relations, one is
led to wonder whether they are valid even when only a {\it minimal} number
of assumptions on the EoS and on how neutron stars are modeled is made;
if so, we can then expect them to be valid for {\it any} physically
reasonable EoS.
This leads to the question of whether one can construct a
reasonable set of minimal assumptions that leads to a family of physically
reasonable EoSs.
Studies along this line were pioneered in the mid-70s by Rhoades and
Ruffini~\cite{Rhoades:1974fn} (see also~\cite{Brecher1976}
and~\cite{1978PhR....46..201H} for a review)
with the goal of obtaining theoretical upper limits on the
mass of neutron stars, important for connecting
massive neutron stars to solar mass black holes and allowing us to distinguish
between these compact objects in x-ray binary observations~\cite{Kalogera:1996ci,Fryer:1999ht}.
This question was partially addressed by~\cite{Steiner:2015aea} who
focused on the $I-\textrm{Love}$ relation (and
other nearly EoS-independent relations, such as that between
the binding energy and the compactness and between the moment of
inertia and compactness).
The goal of this paper is to complete this picture by investigating the
full set of \ilq relations with a physically-reasonable family of EoSs
constructed with minimal assumptions. As an intermediate step, we also
obtain upper bounds on the moment of inertia and quadrupole moment valid
for slowly-rotating neutron stars.

\begin{figure}
\includegraphics[width=1.03\columnwidth]{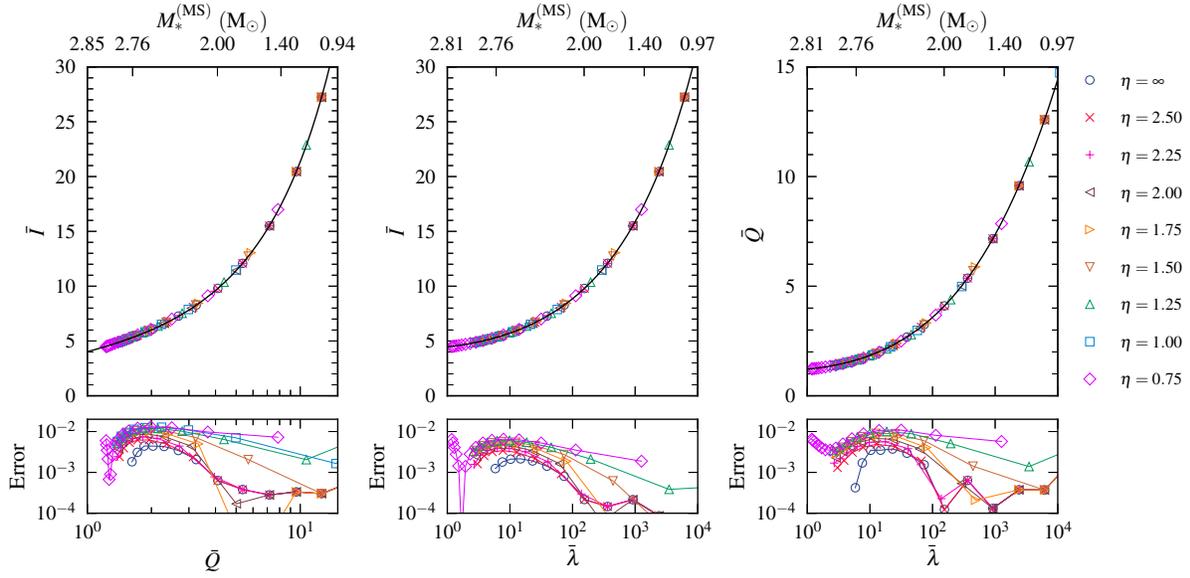}
\caption{The (approximate) EoS-independent ${\bar I}$, ${\bar Q}$
and ${\bar \lambda}$ relations, normalized via ${\bar I} \equiv I / M_{\ast}^3$,
$\bar{\lambda} \equiv \lambda/M_{\ast}^5$ and $\bar{Q} \equiv - Q/(M_{\ast}^3 \chi)$
as defined in~\cite{Yagi:2013bca,Yagi:2013awa},
where $\chi \equiv J / M_{\ast}^2$, $J$ is the spin angular momentum and
$M_{\ast}$ is the mass of the non-rotating configuration.
From left to right, this figure presents
${\bar I}$ versus ${\bar Q}$, ${\bar I}$ versus ${\bar \lambda}$
and ${\bar Q}$ versus ${\bar \lambda}$, computed with a family of EoSs
with $v_{\rm s} = 1$ and different values of $\eta$. The bottom sub-panels
show the relative error $(|y- y^{\rm fit}|/ y^{\rm fit})$ between any
numerical data set and the analytic fits of~\cite{Yagi:2016bkt}
[cf.~Eq.~\eqref{eq:iloveq}].
For reference, the top tick marks indicate the neutron star mass calculated
in the $\eta \to \infty$ limit.
We see that for the range of $\eta$ explored the \ilq relations remain
robust to better than $1\%$ for all stars with
$1 \lesssim M/M_{\odot} \lesssim 3$.
}
\label{fig:iloveq_causal}
\end{figure}

We find that the \ilq relations continue to hold when using such a
family of EoSs as well as when using tabulated (``realist'') EoSs.
We follow Rhoades and Ruffini~\cite{Rhoades:1974fn} and model the EoS
as a piecewise function that reduces to a given tabulated EoS at low
densities and transitions to a stiff EoS with speed of sound
$v_{\rm s} \in \mathbb{R}_{>0}$ at a transition density
$\veps = \eta\, \veps_{\rm n}$, with $\eta \in \mathbb{R}_{>0}$.
Our main findings are summarized in Fig.~\ref{fig:iloveq_causal},
which shows that the \ilq relations are robust to changes to
the EoS parameter $\eta$ for fixed $v_{\rm s} = 1$ (for a similar
plot where we fix $\eta = 1$ and vary $v_{\rm s}$, see
Fig.~\ref{fig:iloveq_acausal}). Since these EoSs
extremize the bulk stellar properties, our results strongly
suggest the following:
%
% \vspace{0.1cm}
% \begin{center}
% \noindent\fbox{\parbox{13cm}{
\begin{framed}
% \begin{center}
\noindent
The \ilq relations are valid to approximately $1\%$ accuracy for {\it any
physically reasonable EoS provided that}
\begin{enumerate}
\item[(i)] the stars do not rotate maximally,
\item[(ii)] magnetic fields are not extremely strong,
\item[(iii)] and general relativity is valid.
\end{enumerate}
% \end{center}
\end{framed}
% }}
% }}
% \end{center}
% \vspace{0.1cm}
%
Strong magnetic fields, of the size expected in magnetars, would contribute
to the elliptical deformation of the neutron star, thus strongly modifying its
quadrupole moment. Extremely rapid rotation, of magnitude comparable to that
of the fastest millisecond pulsars, will also deteriorate the relations.
Although theories outside of general relativity have so far been shown to
also contain \ilq relations, albeit modified from the general relativity ones,
we cannot discard the possibility \emph{a priori} of exotic theories where
these relations do
not hold.

The remainder of this paper explains how these results are obtained in detail.
In Sec.~\ref{sec:eos}, we introduce the EoS used in~\cite{Rhoades:1974fn} and
the physical requirements we require it satisfy. In Sec.~\ref{sec:slowrot},
we describe how we construct slowly-rotating and tidally deformed neutron
stars. In Sec.~\ref{sec:upperbounds}, we present some novel upper bounds on
the reduced rotational quadrupole moment $Q$ and the moment of inertia
$I$ . Finally, in Sec.~\ref{sec:iloveq}, we confront the \ilq relations
against our numerical integrations and close in Sec.~\ref{sec:conclusions}
by summarizing our main findings and discussing some possible future research.
All throughout, we work in geometric units where $c = 1 = G$.

\section{The equation of state}
\label{sec:eos}

In order to obtain a theoretical upper bound on the maximum mass
$M_{\rm max}$ for neutron stars, Rhoades and Ruffini~\cite{Rhoades:1974fn}
assumed that the star consists of two layers:
\begin{enumerate}
\item[(i)] an outer layer whose contribution to the total
(gravitational) mass $M$ is determined by some known EoS, and
\item[(ii)] an inner core whose contribution to the total mass must be
extremized by performing a variational calculation.
\end{enumerate}
The two layers are matched near the nuclear saturation density,
above which the EoS is poorly understood. This calculation is made
with the following additional assumptions:
\begin{enumerate}
\item[(a)] the structure of a neutron stars is determined by the
general relativistic stellar structure equations, i.e.~the
Tolman-Oppenheimer-Volkoff (TOV) equations~\cite{Harrison:1965};
\label{item:tov}
\item[(b)] the energy density is positive ($\veps > 0$) and related
to pressure by a barotropic (single parameter) EoS, i.e. $p = p(\veps)$;
\item[(c)] $\dd p/\dd \veps \geq 0$, such that the sound
speed $v^2_{\rm s} (\equiv dp/d\veps)$ is positive and matter
is stable against microscopic collapse;
\label{item:stability}
\item[(d)] $v_{\rm s} \leq 1$, i.e. fluid perturbations are causal.
\label{item:causality}
\end{enumerate}
Rhoades and Ruffini~\cite{Rhoades:1974fn} showed that the EoS for
the inner core that maximizes the total mass of a static, spherically
symmetric neutron star is that with $v_{\rm s} = 1$, which
can be written analytically as $p = p_0 + (\veps - \veps_0)$, where
$p_0$ and $\veps_0$ are the pressure and energy
densities at the transition between the two layers respectively. They found that
$M_{\rm max}$ is around $\simeq 3.2$\msun, although this bound
does depend on the values of $\veps_0$ and the EoS used in the
low-density region~\cite{Hartle:1977,1978PhR....46..201H,Kalogera:1996ci}.

Strictly speaking, the Rhoades-Ruffini approach is only valid
for static stars. Numerical results supporting the idea that the
EoS proposed in~\cite{Rhoades:1974fn} also yields the maximum mass
of rapidly uniformly rotating neutron stars was reported
in~\cite{Friedman1987ApJ}.
In the same vein as the variational calculations of Ref.~\cite{Rhoades:1974fn},
but considering slowly-rotating neutron stars (in the sense
made precise in Sec.~\ref{sec:slowrot}), Sabbadini and
Hartle~\cite{Sabbadini:1977} showed
that the moment of inertia is maximized when the inner layer has
constant central density $\veps_{\rm c} \geq \veps_{\rm 0}$. Note
however, that an incompressible fluid has the unrealistic
property of having an infinite speed of sound.
Koranda et al. \cite{Koranda:1996jm} (see also~\cite{Haensel1999A&A}
and references therein) showed that the maximum rotation frequency
is obtained for an EoS, where $v_{\rm s} = 1$ in the inner core,
while the outer layer is maximally soft, i.e. $p = 0$ for
$\veps \geq \veps_0$. Assuming a known EoS in the outer layer
has the effect of decreasing this frequency by a few percent.
More recently, Refs.~\cite{VanOeveren:2017xkv} and~\cite{Moustakidis:2016sab}
provided numerical evidence that the tidal Love number is maximized by
the EoS of Rhoades and Ruffini~\cite{Rhoades:1974fn}. The astute reader
will notice that a study on the maximum quadrupole moment that can be supported
by a neutron star is currently missing.

To be conservative in our study of the extremal bulk properties of
neutron stars, we here consider the following extreme EoS (xEoS) family
\begin{equation}
p(\veps) =
\begin{cases}
p_{\rm ms}(\veps), \qquad\qquad\qquad\,\,\,\, \veps < \veps_0 \\
p_{\rm ms}(\veps_0) + v_{\rm s}^2 (\veps - \veps_0),
\quad\, \veps \geq \veps_0.
\label{eq:eos}
\end{cases}
\end{equation}
where $\veps_0$ is a matching density that separates the neutron star
interior into two regions: one where matter has a speed of sound
$v_{\rm s} \in \mathbb{R}_{>0}$ (when the energy density is less than $\veps_0$)
and another one, $p_{\rm ms}(\veps)$, where matter is described by the EoS
MS~\cite{Mueller:1996pm}\footnote{There is some persistent confusion in
the literature regarding the acronym used for this EOS. The MS EoS that
supports neutron stars with maximum masses of $\simeq 2.7$\msun has been
both called MS0 and MS1. To avoid confusing the reader further, we will refer
to this EoS as MS only.}. We thus
follow~\cite{VanOeveren:2017xkv} and assume a relatively stiff
EoS for the known, outer region of the star. We choose the matching
energy density $\veps_0$ to be a multiple of the nuclear
saturation density (taken as $\veps_{\rm n} = 2.7 \times 10^{14}$~\gcmc):
\begin{equation}
\veps_0 \equiv \eta \,\veps_{\rm n},
\label{eq:match}
\end{equation}
where $\eta \in \mathbb{R}_{>0}$. In the limit $\eta \to \infty$,
the xEoS reduces to $p_{\rm ms}(\veps)$,
while as $\eta \to 0$, the EoS is given by
$p = p_{\rm ms}(\veps_0) + v_{\rm s}^2 (\veps - \veps_0)$.
Our choice of EoS differs from that of~\cite{VanOeveren:2017xkv} in
that we allow for $v_{\rm s} \neq 1$ so that we can explore possible
further deviations from universality in the \ilq relations.

The speed of sound controls the stiffness of the xEoS, and thus,
its range warrants some further comments.
On general grounds, an upper bound of $v_{\rm s}^2 = 1/3$ can be
obtained for systems displaying conformal symmetry, which have zero trace
of the energy-momentum tensor.
Calculations of the speed of sound in strongly interacting relativistic
systems (which are not conformal), have shown consistently that
$v_{\rm s}^2 < 1/3$ (see Ref.~\cite{Bedaque:2014sqa} for further details).
Interestingly, Bedaque and Steiner~\cite{Bedaque:2014sqa} found that
the observations of $\simeq 2$\msun neutron
stars~\cite{Demorest:2010bx,Antoniadis:2013pzd}
are in tension with the upper bound $v_{\rm s}^2 = 1/3$,
assuming an EoS of the form of Eq.~\eqref{eq:eos}
with $v_{\rm s}^2 = 1/3$ and with the region below twice
nuclear saturation density
described by a non-relativistic model for hadronic interactions
subject to constrains coming from nuclear physics experiments,
More recently, Alsing et al.~\cite{Alsing2017} found strong statistical
evidence that the maximum speed of sound has a lower bound of
$v_{\rm s} \gtrsim 2/5$ by analyzing the neutron star mass
distribution.

But what about the upper bound of $v_{\rm s}$?
Formally, $\sqrt{\dd p/\dd \veps}$ is the phase
velocity of sound waves in the neutron star fluid. In non-dispersive
fluids, this number coincides with the group velocity, which is
required to be $< 1$ by causality. Neutron star interiors, however,
are expected to be dispersive (see
e.g.~\cite{Friedman1987ApJ} and references therein), and one could
then have a violation of $v_{\rm s} < 1$.
Nonetheless, van Oeveren and Friedman~\cite{VanOeveren:2017xkv}
(based on~\cite{Geroch:1990bw})
have shown recently that causality actually does imply
$\dd p/\dd \veps \leq 1$ for a two parameter EoS, $p=p(\veps, s)$ where
$s$ is the entropy per baryon for stable relativistic fluids.
For more complicated multi-parameter EoSs (e.g. including different
particle species), $\dd p/\dd \veps \leq 1$ follows from local
stability assuming $v_{\rm s} < 1$.
Given all of this, we will here mostly assume that
$v_{\rm s} \leq 1$, but we will consider violations of this conditions
just to see if the \ilq relations continue to hold even then.

To confirm that the xEoS maximizes the values of
the bulk properties of neutron stars, we also consider a few other
less stiff choices of outer EoSs (in comparison with the MS EoS) in the
$\veps < \veps_0$ region. In particular, we will consider the following
outer EoSs:
MPA1~\cite{Muther:1987xaa}, BSk21~\cite{Potekhin:2013qqa} and
SLy4~\cite{Douchin:2001sv}, in decreasing order of stiffness.
All of these support $2$\msun neutron stars as shown in the
left panel of Fig.~\ref{fig:massradius_tp}.

\section{Slowly-rotating and tidally deformed neutron stars}
\label{sec:slowrot}

We construct families of slowly-rotating neutron stars solutions
using the perturbative approach introduced by Hartle
and Thorne~\cite{Hartle:1967he,Hartle:1968ht}, using the xEoS
described in Sec.~\ref{sec:eos}. In this approach, rotation is taken as a
small perturbation upon a static spherically symmetric stellar
background configuration -- a solution of the TOV
equations~\cite{Harrison:1965}. The perturbative
parameter is $\xi \equiv \Omega / \Omega^{\ast}$, where $\Omega$ is the
rotation frequency of the star and $\Omega_{\ast}$ is the (Newtonian)
mass-shedding frequency $\Omega_{\ast} \equiv \sqrt{M_{\ast} / R_{\ast}^3}$,
where in turn $M_{\ast}$ and $R_{\ast}$ are the mass and areal radius
of the non-rotating model.
% WHY USE THE SLOW-ROTATION IS JUSTIFIED:
In realistic astrophysical scenarios, the
slow-rotation approximation is sufficiently accurate. Even
for the fastest spinning neutron star observed,
PSR J1748-2446ad~\cite{Hessels:2006ze}, for which
$\Omega \simeq 4500$ Hz, then $\xi \simeq 0.1$~\cite{Haensel:2007yy}
if we assume it has a mass and radius of $M = 1.4$\msun and $R = 10$~km.

From these families of slowly-rotating neutron stars, we extract their
moment of inertia $I$, the (rotational) quadrupole moment $Q$ and the Love
number. To do so, we begin by integrating the Hartle-Thorne system of equations
at zeroth, first and second perturbative order in $\xi$, in the form used
in~\cite{Sumiyoshi:1999} after correcting some misprints pointed out
in~\cite{Berti:2004ny}.
To test the accuracy of our code, we compare our results to the
tables in~\cite{Berti:2004ny} finding excellent agreement, and we test
our implementation of the xEoS by reproducing the results of Kalogera
and Baym~\cite{Kalogera:1996ci} in the non-rotating limit.
From these numerical solutions, we can then easily extract the moment of inertia
and the quadrupole moment as described, e.g. in~\cite{Yagi:2013awa}.

When present in a binary, a neutron star is tidally deformed by the
gravitational field of its companion. This deformation is predominantly encoded in
the $\ell = 2$ (electric-type) tidal Love number
$\lambda$~\cite{Flanagan:2007ix,Hinderer:2009ca,Binnington:2009bb,Damour:2012yf,Vines:2011ud}.
We calculate this quantity using the formalism of the tidal deformations
of  static neutron star (solutions of the TOV equations), as presented
in~\cite{Postnikov:2010yn}. We validate our numerical calculations
through comparisons with the results of~\cite{Yagi:2013awa}.

\begin{figure}
\centering
\includegraphics[width=0.6\columnwidth]{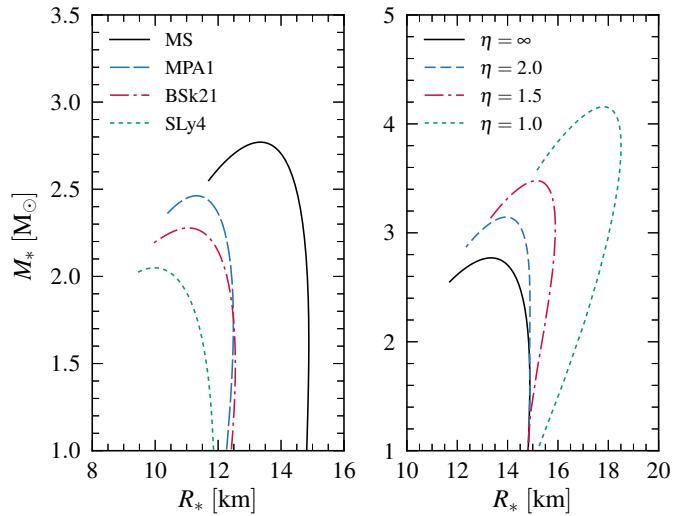}
\caption{Mass-radius relations for non-rotating neutron stars using
a few realistic EoSs (left) and using the xEoS with an outer MS EoS
and various choices of $\eta$ with $v_{\rm s} = 1$ (right).
For $\eta \neq 0$, the
mass-radius relations branch off the $\eta = 0$ curve when the
energy density $\veps_{\rm c}$ at the center of the star is larger
than $\veps_0$.
Note that the axes ranges are drastically different in the left and right panels.}
\label{fig:massradius_tp}
\end{figure}

Figure~\ref{fig:massradius_tp} shows the mass-radius ($M$--$R$) curves for
a family of neutron stars parametrized by central density along any curve,
using a variety of EoSs. The left panel shows
the $M$--$R$ relation using a few different realistic EoSs that span the
entire stellar interior. The right panel shows the same relation but using
xEoS for fixed $v_{\rm s} = 1$ but different choices of the transition density,
as parameterized by $\eta$.
Whenever the central energy density is such
that $\veps_{\rm c} > \veps_0$ (for a given value of $\eta$) the star
has a core region described by a maximally stiff fluid. As a result,
neutron stars then have masses and radii that are substantially
different from that of a star with same central energy density
but described entirely by the realistic EoS.

\section{Upper bounds on the moment of inertia and on the quadrupole moment}
\label{sec:upperbounds}

Before studying the \ilq relations, let us discuss upper bounds
on the moment of inertia and the quadrupole moment. Although similar
bounds for the moment of inertia had already been studied for an
incompressible fluid core in~\cite{Sabbadini:1977,Kalogera:1999nj,Raithel:2016vtt},
ours is,  to our knowledge, the first study of its kind that refers to
the quadrupole moment. Throughout this section, we assume that
$v_{\rm s} = 1$ and only study the sensitivity of the
upper bounds on the transition density $\veps_0$.

In what follows, we will study the behavior of the dimensionless
$\bar{I}$, $\bar{\lambda}$ and $\bar{Q}$, so let us discuss here how
these quantities are defined. The dimensionless moment of inertia
${\bar I} \equiv I / M_{\ast}^3$, the Love number
$\bar{\lambda} \equiv \lambda/M_{\ast}^5$, and the quadrupole
moment $\bar{Q} \equiv - Q/(M_{\ast}^3 \chi)$, where
$\chi \equiv S / M_{\ast}^2$, $S$ is the spin angular momentum
and $Q$ is the quadrupole moment.
As defined here then, ${\bar I}$ and ${\bar Q}$ are independent of
the rotation frequency of the neutron star in the slow-rotation
approximation~\cite{Yagi:2013bca,Yagi:2013awa}.

\begin{figure}
\centering
\includegraphics[width=0.6\columnwidth]{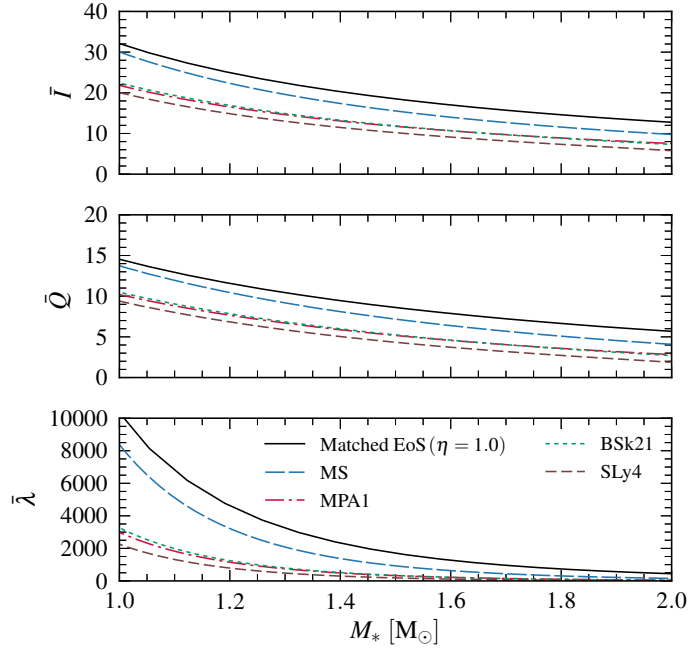}
\caption{The dimensionless quantities ${\bar I}$, ${\bar Q}$ and
${\bar \lambda}$ as a function of the gravitational mass $M_{\ast}$. For a
given mass $M_{\ast}$, the xEoS gives an upper bound for each of these
quantities.
}
\label{fig:reducedquantities_mass}
\end{figure}

\begin{figure}
\centering
\includegraphics[width=0.6\columnwidth]{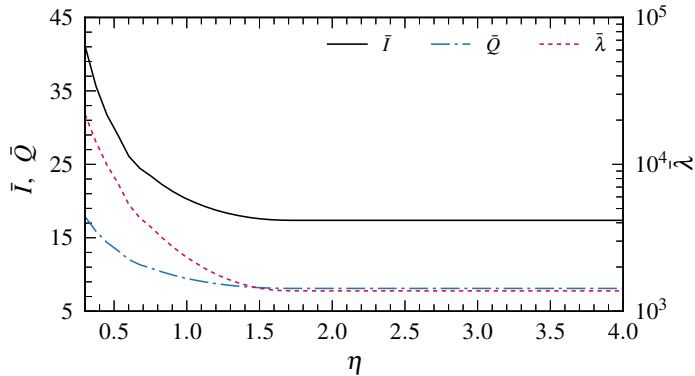}
\caption{
The dimensionless quantities ${\bar I}$, ${\bar Q}$ and ${\bar \lambda}$
(solid, dashed and dotted lines respectively) as a function of $\eta$. We
consider a family of solutions with fixed gravitational mass
$M_{\ast} = 1.4$\msun. In the limit $\eta \to \infty$ the curves converge
to the result of using only EoS MS.}
\label{fig:reducedquantities_eta}
\end{figure}

We start by verifying that the xEoS does indeed maximize
${\bar I}$, ${\bar Q}$ and ${\bar \lambda}$. This is done
Figure~\ref{fig:reducedquantities_mass} shows that the values
of each of these quantities is larger than those obtained when using the
realistic EoS in our catalog. In each panel, we fix $\eta = 1$ in the
xEoS. If we were to increase $\eta$, the curve would
move downwards until, unsurprisingly, it would overlap the MS EoS curve.
On the other hand, if we were to decrease $\eta$, the curve would move
upwards, because then it would begin to resemble an isothermal
fluid sphere (i.e.~a polytope in the $n\to \infty$ limit),
where the universality deteriorates.
This behavior can be observed in Fig.~\ref{fig:reducedquantities_eta},
where we construct families of constant $M_{\ast} = 1.4 {\,\rm M}_{\odot}$
stars using the xEoS for different values of $\eta$. In all cases, ${\bar I}$,
${\bar Q}$ and ${\bar \lambda}$ increase as $\eta \to 0$, while they converge
to the constant values (${\bar I} \simeq 17.4$, ${\bar Q} \simeq 8.10$
and ${\bar \lambda} \simeq 1.37 \times 10^{3}$) as $\eta \to \infty$.


\begin{figure}
\centering
\centering
\includegraphics[width=0.60\columnwidth]{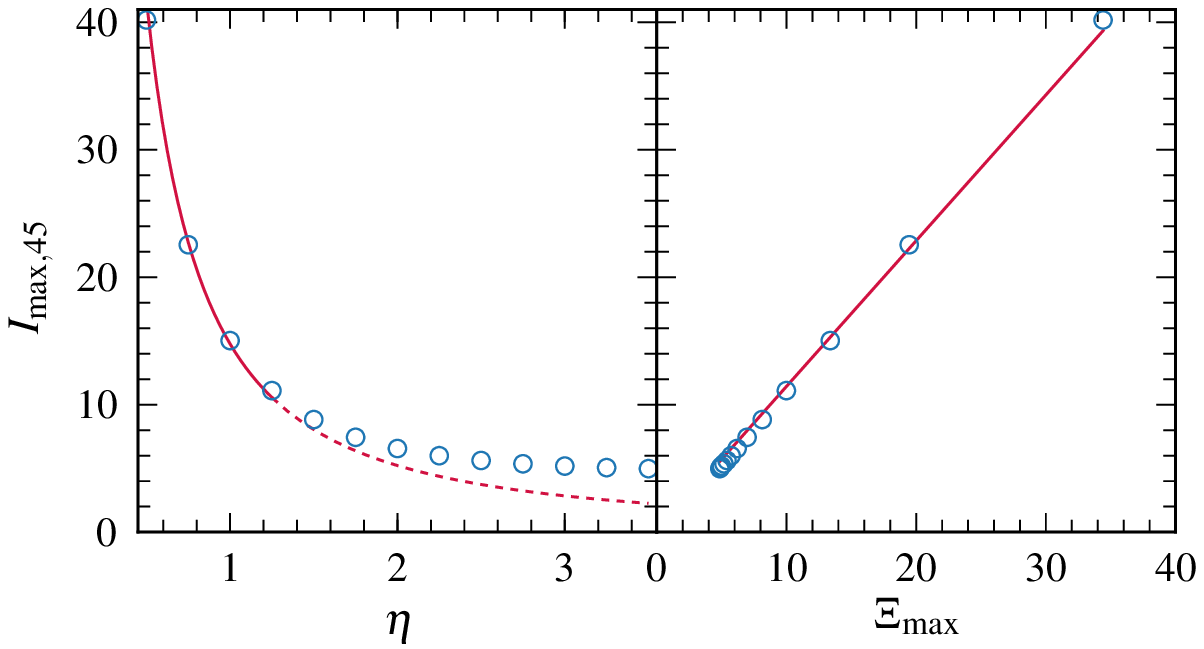} \\
\includegraphics[width=0.60\columnwidth]{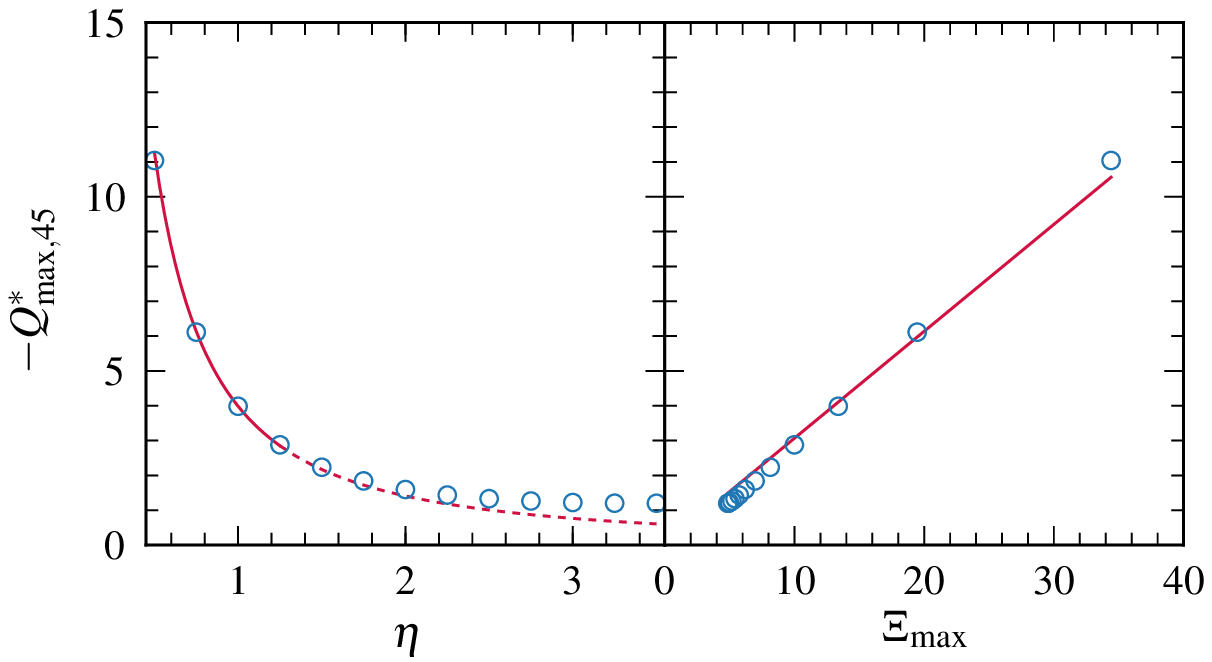}
\caption{The maximum values for the moment of inertia and rotational
quadrupole moment. The top panels show the values of $I_{\rm{max}, 45}$
as functions of $\eta$ (top-left) and $\Xi_{\rm max}$ (top-right), including
the fitting expressions (lines)~\eqref{eq:imax_eta} and~\eqref{eq:imax_xi}.
The bottom panels show that $Q^{\ast}_{\rm{max}, 45}$ has similar behavior,
modulo numerical factors [cf.~Eqs.~\eqref{eq:qmax_eta} and~\eqref{eq:qmax_xi}].
The corresponding curves for $Q_{\rm{max}, 45}$ can easily be obtained
by rescaling it with $\xi^2$. The minus sign in the bottom panels
indicate that the surface of the rotating star
has an oblate geometry. In the left panels, the solid line
represent approximately where the scaling laws~\eqref{eq:imax_eta}
and~\eqref{eq:qmax_eta} are valid ($\eta \lesssim 1.25$),
while the dotted portion of the curves illustrate the breakdown of the
scaling.
}
\label{fig:iq_max}
\end{figure}

Besides examining the dimensionless ${\bar I}$, ${\bar Q}$ and
${\bar \lambda}$, it is also interesting to consider the regular,
dimensionful, moment of inertia and quadrupole moment.
Figure~\ref{fig:iq_max} (upper panels)
shows the maximum moment of inertia
$I_{{\rm{max}}, 45} \equiv I_{\rm{max}} / (10^{45} {\textrm{g cm}}^{2})$
as a function of $\eta$.
As in the case of realistic EoSs, we find that the maximum moment of inertia
occurs for a neutron star with mass
$M \lesssim M_{\rm max}$~\cite{Haensel:2007yy}. This seems to be
in contradiction with the behavior of ${\bar I} = {\bar I} (M_{\ast})$
shown in Fig.~\ref{fig:reducedquantities_mass}; in reality it is not
because ${\bar I}$ is normalized by $M^{-3}_{\ast}$, which explains
the increase of ${\bar I}$ at lower masses.
For large $\eta$, $I_{{\rm{max}}, 45}$ converges to $\simeq 4.9$,
the maximum moment of inertia we would find using the MS EoS.
In the other limit, $\eta \to 0$, the only scaling parameter is the transition
density $\veps_0$ which has units of length$^{-2}$. Since the moment of
inertia has units of length$^3$, we find it scales as $\veps_0^{-3/2}$.
More precisely,
\begin{equation}
I_{\rm{max}, 45} \simeq 14.7 \times
\left(\frac{\veps_0}{\veps_{\rm n}}\right)^{-3/2}, \qquad \eta \lesssim 1.25
\label{eq:imax_eta}
\end{equation}
which is shown by the solid line in the top-left panel of
Fig.~\ref{fig:iq_max}. This same reasoning has been used to explain the
low $\eta$ behavior of the maximum mass, radius and tidal Love number
in~\cite{Koranda:1996jm,VanOeveren:2017xkv}.

Let us now consider the relation between $I_{\rm{max}}$ and
the maximum mass $M_{\rm max}$ and its radius $R_{M_{\rm max}}$. For
realistic EoSs, these quantities have been numerically found to be
related by
\begin{equation}
I_{\rm{max}, 45} \simeq
k \, \Xi_{\rm max}\,, \qquad
\Xi_{\rm max} \equiv \left(\frac{M_{\rm max}}{{\rm M}_{\odot}}\right)
\left(\frac{R_{M_{\rm{max}}}}{10\,{\rm km}}\right)^2,
\label{eq:imax_xi}
\end{equation}
where $k \approx 0.97$~\cite{Bejger:2002ty}\footnote{An improved
fit can be obtained by also adding the compactness $C_{\rm max}\equiv
({M_{\rm max}}/{{\rm M}_{\odot}})/(R_{M_{\rm max}}/{\rm km}$) in the
fitting function (see e.g.~Eq.~(11) in \cite{Bejger:2005jy}).}. Using
the xEoS, we find a similar relationship, but with
$k = 1.14$, as shown in the top-right panel of Fig.~\ref{fig:iq_max}.

The maximum quadrupole moment obeys relations similar to those
for the moment of inertia, although a note of warning is necessary.
The quadrupole moment $Q$, unlike the moment of inertia $I$,
scales with the expansion parameter
$\xi$ as $Q = \xi^2 Q_{\ast}$, where $Q_{\ast}$ is
the quadrupole moment of a star rotating at the mass-shedding frequency
$\Omega_{\ast}$. Since by definition $0 \leq \xi \leq 1$, the
most conservative upper bound on the quadrupole
moment would seem to be $Q < Q_{\ast}$, but this is incorrect because
the $Q = \xi^2 Q_{\ast}$ scaling is only valid in the slow-rotation
approximation.
Comparing the calculation of $Q$
using the Hartle-Thorne formalism to full numerical integrations
using the RNS code~\cite{Stergioulas:1994ea}, Berti et al.~\cite{Berti:2004ny}
found that the slow-rotation and exact $Q$ disagree by $10\%$ for
stellar models with $\xi \approx 0.2$ when $M = 1.4$\msun and by $20\%$
for the maximum mass configuration (for a given EoS). These errors grow
with increasing $\xi$, with the Hartle-Thorne calculation systematically
overestimating the values of $Q$.
%
% Nonetheless, in realistic astrophysical scenarios, the
% slow-rotation approximation is sufficiently accurate. Even
% for the fastest spinning neutron star,
% PSR J1748-2446ad~\cite{Hessels:2006ze}, for which
% $\Omega \simeq 4500$ Hz, if we assume it to have canonical mass
% and radius $M = 1.4$\msun and $R = 10$~km gives
% $\xi \simeq 0.1$~\cite{Haensel:2007yy}.

Akin to the case of $I_{\rm{max}}$, we find that the
maximum quadrupole moment
[$Q^{\ast}_{\rm{max}, 45} \equiv Q^{\ast}_{\rm{max}} / (10^{45} \textrm{ g cm}^2)$]
satisfies
\begin{equation}
Q_{\rm{max}, 45} \equiv \xi^2 Q^{\ast}_{\rm{max}, 45}
\simeq 0.31\,\xi^2 \times
\left(\frac{\veps_0}{\veps_{\rm n}}\right)^{-3/2}, \qquad \eta \lesssim 1.25
\label{eq:qmax_eta}
\end{equation}
and
\begin{equation}
Q_{\rm{max}, 45} =
\xi^2 Q^{\ast}_{\rm{max}, 45}
\simeq
3.98\,\xi^2 \, \Xi_{\rm max}\,,
\label{eq:qmax_xi}
\end{equation}
as shown in the bottom panels of Fig.~\ref{fig:iq_max}.
We expect these relations to hold for neutron stars spinning
up to $\xi \lesssim 0.1$, based on~\cite{Berti:2004ny}.
To extend the applicability of Eqs.~\eqref{eq:qmax_eta}
and~\eqref{eq:qmax_xi} to $\xi \gtrsim 0.1$, a study
considering rapidly rotating neutron stars would be
required~\cite{Paschalidis:2016vmz}, but this is outside the
scope of this paper.

\section{Validity of the \ilq relations for
the matched equation of state}
\label{sec:iloveq}

\subsection{Universal relations and the matched equation of state}

Let us now test the \ilq relations using the xEoS of Eq.~\eqref{eq:eos}.
In previous work, Yagi and Yunes~\cite{Yagi:2013bca,Yagi:2013awa} showed
that the \ilq relations can be fitted to the function
\begin{equation}
\ln y_{i} = \sum_{k=0}^{4} c_{k}( \ln x_{i})^k,
\label{eq:iloveq}
\end{equation}
where $y_i, x_i$ are a pair of variables from the ${\bar I}$, ${\bar Q}$,
${\bar \lambda}$ trio and $c_k$ are numerical (fitting) constants (see
Table 1 in~\cite{Yagi:2016bkt}). This fit was carried out using a very large
sample of tabulated (realistic) EoSs, but of course nobody has
yet considered the \ilq relations for the xEoS of Eq.~\eqref{eq:eos},
as we do next in this section.

Figure~\ref{fig:iloveq_causal} already showed that the \ilq relations
are satisfied for the xEoS of Eq.~\eqref{eq:eos}, using a variety of
transition energy densities $\veps_{0}$ parameterized by $\eta$ with a
fixed sound speed $v_{\rm s} = 1$. Section~\ref{sec:slowrot} taught us
that the lower the value of $\eta$, the larger the values of ${\bar I}$,
${\bar Q}$ and ${\bar \lambda}$ (cf.~Fig.~\ref{fig:reducedquantities_eta}).
As shown in the bottom panels of Fig.~\ref{fig:iloveq_causal}, the
non-universality of the \ilq relations (as quantified
by the relative fractional error from the fits) remain always
below $\simeq 1\%$ for the range of $\eta$ we considered.

\begin{figure*}
\includegraphics[width=1.03\columnwidth]{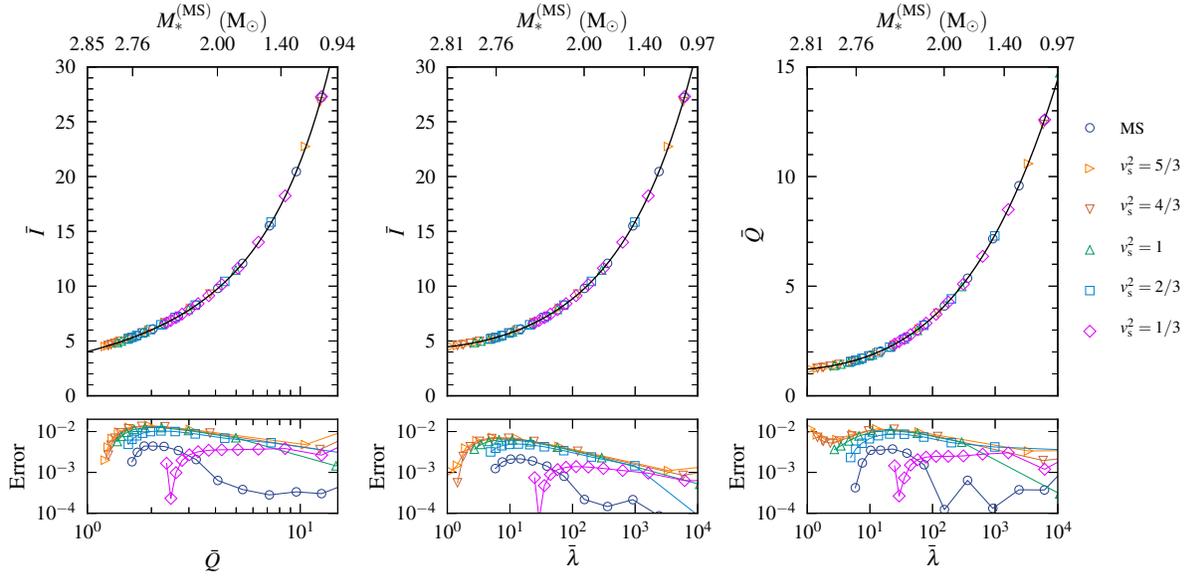}
\caption{\ilq relations for fixed transition density with $\eta = 1$ but varying
speed of sound $v_{\rm s}$. As in the case of Fig.~\ref{fig:iloveq_causal},
the \ilq relations are EoS-independent to better than $\simeq 1\%$ for all
values of $v_{\rm s}$ considered, including acausal fluids with $v_{\rm s} > 1$.
}
\label{fig:iloveq_acausal}
\end{figure*}

Let us now repeat this study but holding the transition density fixed by
setting $\eta = 1$ and varying the sound speed $v_{\rm s}$.
Figure~\ref{fig:iloveq_acausal} shows the \ilq relations for
$v_{\rm s} \in (1/3,5/3)$, thus including examples of (admittedly unphysical)
superluminal values. Surprisingly, even in this extremal
(and unphysical) situation, the \ilq relations stand firm and once
again the relative fractional errors with respect to the
fitting functions remain below $\simeq 1\%$.

In both Figs.~\ref{fig:iloveq_causal} and~\ref{fig:iloveq_acausal},
we focused on neutron stars with $M \gtrsim 1$\msun, which falls in
the range of the lowest neutron star masses observed~\cite{Lattimer:2012nd},
with the lowest mass precisely measured so far having
$M = 1.174 \pm 0.004$\msun~\cite{Martinez:2015mya}.
For very low masses $M \simeq 0.2$\msun, it is known that the
EoS-universality breaks down~\cite{Yagi:2013awa}
(see also~\cite{Silva:2016myw}). However, the existence of neutron stars
with such low-masses is at odds with supernova studies which
estimate a minimum mass in the $1.15$--$1.20$\msun
range~(cf. Sec. 3.2 of \cite{Lattimer:2012nd}). Low-mass neutron stars
could be formed in speculative scenarios involving the fragmentation
of rapidly rotating proto-neutron stars~\cite{Popov:2006ki}.
It is however sensible to assume $M \geq 1$\msun as a lower bound,
above which we have shown that the \ilq relations hold.

We recall that the xEoS is: (i) devised by
assuming only a small number of physically sensible requirements
on the properties of matter at supranuclear densities and on how stars
are constructed and (ii) as a consequence, it extremizes\footnote{More precisely,
it results in the largest values of ${\bar I}$, ${\bar Q}$ and ${\bar \lambda}$
for a neutron star with fixed mass $M_{\ast}$ in comparison to the other
EoSs in our catalog (cf. Fig.~\ref{fig:reducedquantities_mass}).} the values
of ${\bar I}$, ${\bar Q}$ and ${\bar \lambda}$. The results summarized in
Figs.~\ref{fig:iloveq_causal} and~\ref{fig:iloveq_acausal}, thus provide
{\it strong support for the validity of the \ilq relations for any sensible
EoS that might be used to model neutron stars in general relativity
and for any possible value attainable by these quantities.}

\subsection{Why \ilq so much?}

We have seen that the \ilq relation are robust under rather extreme
variations of the parameters in the xEoS, but why is this so? While the
explanation for the origin of the \ilq remains
elusive, Yagi et al.~\cite{Yagi:2014qua} (see also~\cite{Stein:2013ofa})
suggested that the (nearly) EoS-independence of these relations could be a
result of an emergent symmetry due to an approximate self-similarity of
the ellipsoidal isodensity contours within the star. This symmetry emerges
when we flow from Newtonian, low-compactness ($M_{\ast} / R_{\ast} << 1$)
stars towards relativistic, high-compactness ($M_{\ast} / R_{\ast} \simeq 0.1$)
neutron stars.
In the former case, numerical integration has revealed a breakdown
of both the self-similarity of isodensity contours and
of the $I-Q$ relations~\cite{Yagi:2014qua}.
On the other hand, for neutron stars, Ref.~\cite{Yagi:2014qua} found that
in the region $r \in (0.50 , 0.95) R_{\ast}$
(which contributes the most to the calculation of the moment of inertia and the
quadrupole moment) the eccentricity $e$ of isodensity contours is
approximately constant, changing at most by $\simeq 10\%$. For comparison,
this change can exceed $\simeq 100 \%$ in Newtonian configurations.

\begin{figure}
\centering
\includegraphics[width=0.6\columnwidth]{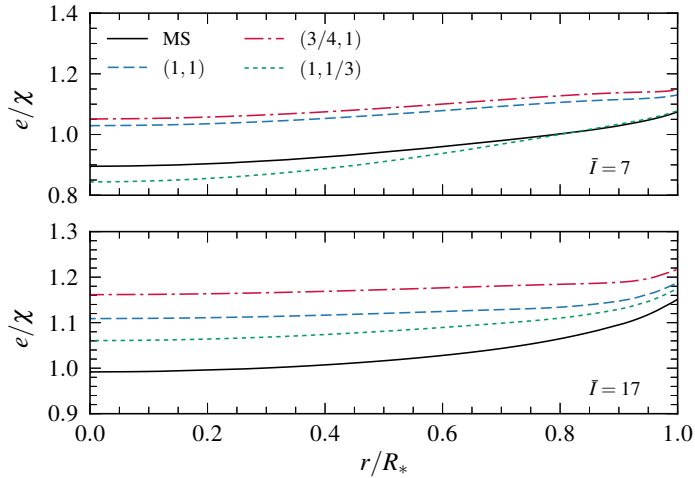}
\caption{Radial profiles of the stellar eccentricity $e$ divided by the spin
parameter $\chi$ for slowly-rotating neutron star configurations with
${\bar I} = 7$ (top) and ${\bar I} = 17$ (bottom). The curves represent
different neutron star models parametrized by ($\eta, v_{\rm s}^2$)
in the xEoS. In the region most relevant for the calculation of $I$ and $Q$,
$r \in [0.50 R_{\ast}, 0.95 R_{\ast}]$, the curves vary
at most by $\sim 10\%$. The quantity $e/\chi$ is independent of the spin
rate for slowly-rotating stars.
}
\label{fig:ellip}
\end{figure}

Let us then study whether this self-similarity of the ellipticity
of isodensity layers continues to hold with the xEoS.
Figure~\ref{fig:ellip} shows the ellipticity profiles, calculated
following~\cite{Hartle:1968ht}, for two neutron star models:
one with ${\bar I} = 7$ ($M \simeq 2.4$\msun using the MS EoS)
and another one with ${\bar I} = 17$ ($M \simeq 1.4$\msun using the MS EoS).
Each panel shows the eccentricity profiles (normalized to
the spin parameter $\chi$) for four pairs of values of
$(\eta, v^2_{\rm s})$. In all cases, we see that $e/\chi$ changes
very mildly in the region $r \in (0.50 , 0.95) R_{\ast}$,
just as~\cite{Yagi:2014qua} found in the case of realistic EoSs.
In fact, in some cases, such as when ${\bar I} = 17$ and $(\eta, v_{\rm s}^2)
= (3/4, 1)$, the eccentricity profile becomes almost flat, except near
$r \gtrsim 0.95 R_{\ast}$.
These results strengthen the case that the approximately constant
ellipticity of isodensity levels could be an explanation for the \ilq relations.

\section{Conclusions and Outlook}
\label{sec:conclusions}

The \ilq relations, provide an unique way of breaking the EoS-degeneracy
by establishing EoS-independent relations between the moment of inertia,
the quadrupole moment and the tidal Love number. This feature, makes them
powerful tools for breaking degeneracies in various astrophysical
situations, as in pulse profile modeling of rotating neutron
stars~\cite{Baubock:2013gna} with immediate impact to current
(such as NICER~\cite{Gendreau2012,Arzoumanian2014SPIE}) and future
observatories (such as eXTP~\cite{Zhang:2016ach}).
In the gravitational wave astronomy arena, the \ilq relations can
break degeneracies in the measurement of the tidal deformability in merger
events involving neutron stars.
Due to their potential usefulness, it is natural to ask whether the
\ilq relations remain valid when we demand only a limited, physically
sound, number of requirement on the EoS above
nuclear saturation density.
We have shown that the \ilq relations are surprisingly robust
against an agnostic EoS model, putting them on firm ground
for astrophysical applications, despite EoS uncertainties.

Our work can be extended in a number of directions. The most immediate
direction would be to consider rapidly rotating neutron
stars~\cite{Doneva:2013rha,Chakrabarti:2013tca}
and to explore the validity of the three-hair
relations~\cite{Pappas:2013naa,Yagi:2014bxa} using the xEoS.
Alternatively, one could consider higher-order calculations
within the Hartle-Thorne perturbative scheme.
Rotating neutron star models at third-order~\cite{Benhar:2005gi} and
fourth-order~\cite{Yagi:2014bxa} in the slow-rotation expansion
have indeed been obtained in the past.

Another interesting avenue for future research is to investigate the
maximum value of the bulk properties of neutron stars in modified theories
of gravity (see Refs.~\cite{Clifton:2011jh,Yunes:2013dva,Berti:2015itd}
for reviews).
One of the major difficulties in using neutron stars
as tests of general relativity is the degeneracy between our ignorance
on the EoS and modifications to Einstein's
gravity~\cite{Glampedakis:2015sua,Glampedakis:2016pes}.
If we assume the same xEoS to obtain upper bounds
on various bulk properties of neutron stars, either on
a gravity-by-gravity theory case or in a parametrized
way~\cite{Glampedakis:2015sua}, we can potentially use future
observations to signal the presence of new physics.
Such upper bounds (notably on the maximum mass) were
obtained in the past (cf.~\cite{1978PhR....46..201H} for a review
on early works in this direction). However, these works studied
theories that are of little interest today (exceptions include
Einstein-dilaton-Gauss-Bonnet gravity~\cite{Pani:2011xm}
and scalar-tensor gravity~\cite{Zaglauer:1992bp,Sotani:2017pfj}).
It would be interesting to revisit this problem using more modern
alternatives to general relativity.

\section*{Acknowledgements}

H.O.S thanks Thomas Sotiriou and the University of Nottingham
for the support and warm hospitality during the initial stages of
this work. He also thanks Davide Gerosa, for useful advice on
{\sc matplotlib}~\cite{Hunter:2007} used to generate the figures and
Leandro A. Oliveira for discussions on dispersive fluids.
H.O.S and N.Y thank Kent Yagi for sharing some equations of state
tables for this work, validating some of our numerical
calculations, profitable discussions and comments on this
paper.~H.O.S and N.Y acknowledge financial support through NSF CAREER
grant PHY-1250636 and NASA grants NNX16AB98G and 80NSSC17M0041.
%%%%%%%%%%%%%%%%%%%%%%%%%%%%%%%%%%%%%%%%%%%%%%%%%%

%%%%%%%%%%%%%%%%%%%%%%%%%%%%%%%%%%%%%%%%%%%%
\section*{References}
\bibliographystyle{iopart-num}
% \bibliography{biblio}

\begin{thebibliography}{10}
\expandafter\ifx\csname url\endcsname\relax
  \def\url#1{{\tt #1}}\fi
\expandafter\ifx\csname urlprefix\endcsname\relax\def\urlprefix{URL }\fi
\providecommand{\eprint}[2][]{\href{http://arxiv.org/abs/#2}{arXiv:#2}}

\bibitem{Lattimer:2012nd}
Lattimer J~M 2012 {\em Ann. Rev. Nucl. Part. Sci.\/} {\bf 62} 485--515
  [\eprint{1305.3510}]

\bibitem{Miller:2016pom}
Miller M~C and Lamb F~K 2016 {\em Eur. Phys. J.\/} {\bf A52} 63
  [\eprint{1604.03894}]

\bibitem{Watts:2016uzu}
Watts A~L {\em et~al.\/} 2016 {\em Rev. Mod. Phys.\/} {\bf 88} 021001
  [\eprint{1602.01081}]

\bibitem{Demorest:2010bx}
Demorest P, Pennucci T, Ransom S, Roberts M and Hessels J 2010 {\em Nature\/}
  {\bf 467} 1081--1083 [\eprint{1010.5788}]

\bibitem{Antoniadis:2013pzd}
Antoniadis J {\em et~al.\/} 2013 {\em Science\/} {\bf 340} 6131
  [\eprint{1304.6875}]

\bibitem{Antoniadis:2016hxz}
Antoniadis J, Tauris T~M, Özel F, Barr E, Champion D~J and Freire P~C~C 2016
  [\eprint{1605.01665}]

\bibitem{Alsing2017}
Alsing J, Silva H~O and Berti E 2017  [\eprint{1709.07889}]

\bibitem{Ozel:2016oaf}
Özel F and Freire P 2016 {\em Ann. Rev. Astron. Astrophys.\/} {\bf 54} 401
  [\eprint{1603.02698}]

\bibitem{Barker1975}
{Barker} B~M and {O'Connell} R~F 1975 {\em \prd\/} {\bf 12} 329--335

\bibitem{Damour1988}
{Damour} T and {Schafer} G 1988 {\em Nuovo Cimento B Serie\/} {\bf 101}
  127--176

\bibitem{Burgay:2003jj}
Burgay M {\em et~al.\/} 2003 {\em Nature\/} {\bf 426} 531--533
  [\eprint{astro-ph/0312071}]

\bibitem{Lyne:2004cj}
Lyne A~G {\em et~al.\/} 2004 {\em Science\/} {\bf 303} 1153--1157
  [\eprint{astro-ph/0401086}]

\bibitem{Lattimer:2004nj}
Lattimer J~M and Schutz B~F 2005 {\em Astrophys. J.\/} {\bf 629} 979--984
  [\eprint{astro-ph/0411470}]

\bibitem{Cadeau:2004gm}
Cadeau C, Leahy D~A and Morsink S~M 2005 {\em Astrophys. J.\/} {\bf 618}
  451--462 [\eprint{astro-ph/0409261}]

\bibitem{Cadeau:2006dc}
Cadeau C, Morsink S~M, Leahy D and Campbell S~S 2007 {\em Astrophys. J.\/} {\bf
  654} 458--469 [\eprint{astro-ph/0609325}]

\bibitem{Morsink:2007tv}
Morsink S~M, Leahy D~A, Cadeau C and Braga J 2007 {\em Astrophys. J.\/} {\bf
  663} 1244--1251 [\eprint{astro-ph/0703123}]

\bibitem{Psaltis:2013fha}
Psaltis D, Özel F and Chakrabarty D 2014 {\em Astrophys. J.\/} {\bf 787} 136
  [\eprint{1311.1571}]

\bibitem{Psaltis:2013zja}
Psaltis D and Özel F 2014 {\em Astrophys. J.\/} {\bf 792} 87
  [\eprint{1305.6615}]

\bibitem{Mora:2003wt}
Mora T and Will C~M 2004 {\em Phys. Rev.\/} {\bf D69} 104021 [Erratum: Phys.
  Rev.D71,129901(2005)] [\eprint{gr-qc/0312082}]

\bibitem{Berti:2007cd}
Berti E, Iyer S and Will C~M 2008 {\em Phys. Rev.\/} {\bf D77} 024019
  [\eprint{0709.2589}]

\bibitem{Flanagan:2007ix}
Flanagan E~E and Hinderer T 2008 {\em Phys. Rev.\/} {\bf D77} 021502
  [\eprint{0709.1915}]

\bibitem{Read:2009yp}
Read J~S, Markakis C, Shibata M, Uryu K, Creighton J~D~E and Friedman J~L 2009
  {\em Phys. Rev.\/} {\bf D79} 124033 [\eprint{0901.3258}]

\bibitem{Read:2013zra}
Read J~S, Baiotti L, Creighton J~D~E, Friedman J~L, Giacomazzo B, Kyutoku K,
  Markakis C, Rezzolla L, Shibata M and Taniguchi K 2013 {\em Phys. Rev.\/}
  {\bf D88} 044042 [\eprint{1306.4065}]

\bibitem{Lackey:2013axa}
Lackey B~D, Kyutoku K, Shibata M, Brady P~R and Friedman J~L 2014 {\em Phys.
  Rev.\/} {\bf D89} 043009 [\eprint{1303.6298}]

\bibitem{Yagi:2015pkc}
Yagi K and Yunes N 2016 {\em Class. Quant. Grav.\/} {\bf 33} 13LT01
  [\eprint{1512.02639}]

\bibitem{Dietrich:2015pxa}
Dietrich T, Moldenhauer N, Johnson-McDaniel N~K, Bernuzzi S, Markakis C~M,
  Brügmann B and Tichy W 2015 {\em Phys. Rev.\/} {\bf D92} 124007
  [\eprint{1507.07100}]

\bibitem{DelPozzo:2013ala}
Del~Pozzo W, Li T~G~F, Agathos M, Van Den~Broeck C and Vitale S 2013 {\em Phys.
  Rev. Lett.\/} {\bf 111} 071101 [\eprint{1307.8338}]

\bibitem{Andersson:1996pn}
Andersson N and Kokkotas K~D 1996 {\em Phys. Rev. Lett.\/} {\bf 77} 4134--4137
  [\eprint{gr-qc/9610035}]

\bibitem{Andersson:1997rn}
Andersson N and Kokkotas K~D 1998 {\em Mon. Not. Roy. Astron. Soc.\/} {\bf 299}
  1059--1068 [\eprint{gr-qc/9711088}]

\bibitem{Yagi:2013bca}
Yagi K and Yunes N 2013 {\em Science\/} {\bf 341} 365--368 [\eprint{1302.4499}]

\bibitem{Yagi:2013awa}
Yagi K and Yunes N 2013 {\em Phys. Rev.\/} {\bf D88} 023009
  [\eprint{1303.1528}]

\bibitem{Yagi:2016bkt}
Yagi K and Yunes N 2017 {\em Phys. Rept.\/} {\bf 681} 1--72
  [\eprint{1608.02582}]

\bibitem{Doneva:2017jop}
Doneva D~D and Pappas G 2017  [\eprint{1709.08046}]

\bibitem{Rhoades:1974fn}
Rhoades Jr C~E and Ruffini R 1974 {\em Phys. Rev. Lett.\/} {\bf 32} 324--327

\bibitem{Brecher1976}
{Brecher} K and {Caporaso} G 1976 {\em \nat\/} {\bf 259} 377

\bibitem{1978PhR....46..201H}
{Hartle} J~B 1978 {\em \physrep\/} {\bf 46} 201--247

\bibitem{Kalogera:1996ci}
Kalogera V and Baym G 1996 {\em Astrophys. J.\/} {\bf 470} L61--L64
  [\eprint{astro-ph/9608059}]

\bibitem{Fryer:1999ht}
Fryer C~L and Kalogera V 2001 {\em Astrophys. J.\/} {\bf 554} 548--560
  [\eprint{astro-ph/9911312}]

\bibitem{Steiner:2015aea}
Steiner A~W, Lattimer J~M and Brown E~F 2016 {\em Eur. Phys. J.\/} {\bf A52} 18
  [\eprint{1510.07515}]

\bibitem{Harrison:1965}
{Harrison} B~K, {Thorne} K~S, {Wakano} M and {Wheeler} J~A 1965 {\em
  {Gravitation Theory and Gravitational Collapse}\/}

\bibitem{Hartle:1977}
{Hartle} J~B and {Sabbadini} A~G 1977 {\em \apj\/} {\bf 213} 831--835

\bibitem{Friedman1987ApJ}
{Friedman} J~L and {Ipser} J~R 1987 {\em \apj\/} {\bf 314} 594--597

\bibitem{Sabbadini:1977}
{Sabbadini} A~G and {Hartle} J~B 1977 {\em Annals of Physics\/} {\bf 104}
  95--133

\bibitem{Koranda:1996jm}
Koranda S, Stergioulas N and Friedman J~L 1997 {\em Astrophys. J.\/} {\bf 488}
  799 [\eprint{astro-ph/9608179}]

\bibitem{Haensel1999A&A}
{Haensel} P, {Lasota} J~P and {Zdunik} J~L 1999 {\em \aap\/} {\bf 344} 151--153

\bibitem{VanOeveren:2017xkv}
Van~Oeveren E~D and Friedman J~L 2017 {\em Phys. Rev.\/} {\bf D95} 083014
  [\eprint{1701.03797}]

\bibitem{Moustakidis:2016sab}
Moustakidis C~C, Gaitanos T, Margaritis C and Lalazissis G~A 2017 {\em Phys.
  Rev.\/} {\bf C95} 045801 [Erratum: Phys. Rev.C95,no.5,059904(2017)]
  [\eprint{1608.00344}]

\bibitem{Mueller:1996pm}
Muller H and Serot B~D 1996 {\em Nucl. Phys.\/} {\bf A606} 508--537
  [\eprint{nucl-th/9603037}]

\bibitem{Bedaque:2014sqa}
Bedaque P and Steiner A~W 2015 {\em Phys. Rev. Lett.\/} {\bf 114} 031103
  [\eprint{1408.5116}]

\bibitem{Geroch:1990bw}
Geroch R~P and Lindblom L 1990 {\em Phys. Rev.\/} {\bf D41} 1855

\bibitem{Muther:1987xaa}
Muther H, Prakash M and Ainsworth T~L 1987 {\em Phys. Lett.\/} {\bf B199}
  469--474

\bibitem{Potekhin:2013qqa}
Potekhin A~Y, Fantina A~F, Chamel N, Pearson J~M and Goriely S 2013 {\em
  Astron. Astrophys.\/} {\bf 560} A48 [\eprint{1310.0049}]

\bibitem{Douchin:2001sv}
Douchin F and Haensel P 2001 {\em Astron. Astrophys.\/} {\bf 380} 151
  [\eprint{astro-ph/0111092}]

\bibitem{Hartle:1967he}
Hartle J~B 1967 {\em Astrophys. J.\/} {\bf 150} 1005--1029

\bibitem{Hartle:1968ht}
{Hartle} J~B and {Thorne} K~S 1968 {\em \apj\/} {\bf 153} 807

\bibitem{Hessels:2006ze}
Hessels J~W~T, Ransom S~M, Stairs I~H, Freire P~C~C, Kaspi V~M and Camilo F
  2006 {\em Science\/} {\bf 311} 1901--1904 [\eprint{astro-ph/0601337}]

\bibitem{Haensel:2007yy}
Haensel P, Potekhin A~Y and Yakovlev D~G 2007 {\em {Neutron stars 1: Equation
  of state and structure}\/} vol 326 (New York, USA: Springer)

\bibitem{Sumiyoshi:1999}
{Sumiyoshi} K, {Ib{\'a}{\~n}ez} J~M and {Romero} J~V 1999 {\em \aaps\/} {\bf
  134} 39--52

\bibitem{Berti:2004ny}
Berti E, White F, Maniopoulou A and Bruni M 2005 {\em Mon. Not. Roy. Astron.
  Soc.\/} {\bf 358} 923--938 [\eprint{gr-qc/0405146}]

\bibitem{Hinderer:2009ca}
Hinderer T, Lackey B~D, Lang R~N and Read J~S 2010 {\em Phys. Rev.\/} {\bf D81}
  123016 [\eprint{0911.3535}]

\bibitem{Binnington:2009bb}
Binnington T and Poisson E 2009 {\em Phys. Rev.\/} {\bf D80} 084018
  [\eprint{0906.1366}]

\bibitem{Damour:2012yf}
Damour T, Nagar A and Villain L 2012 {\em Phys. Rev.\/} {\bf D85} 123007
  [\eprint{1203.4352}]

\bibitem{Vines:2011ud}
Vines J, Flanagan E~E and Hinderer T 2011 {\em Phys. Rev.\/} {\bf D83} 084051
  [\eprint{1101.1673}]

\bibitem{Postnikov:2010yn}
Postnikov S, Prakash M and Lattimer J~M 2010 {\em Phys. Rev.\/} {\bf D82}
  024016 [\eprint{1004.5098}]

\bibitem{Kalogera:1999nj}
Kalogera V and Psaltis D 2000 {\em Phys. Rev.\/} {\bf D61} 024009
  [\eprint{astro-ph/9903415}]

\bibitem{Raithel:2016vtt}
Raithel C~A, Özel F and Psaltis D 2016 {\em Phys. Rev.\/} {\bf C93} 032801
  [Addendum: Phys. Rev.C93,no.4,049905(2016)] [\eprint{1603.06594}]

\bibitem{Bejger:2002ty}
Bejger M and Haensel P 2002 {\em Astron. Astrophys.\/} {\bf 396} 917
  [\eprint{astro-ph/0209151}]

\bibitem{Bejger:2005jy}
Bejger M, Bulik T and Haensel P 2005 {\em Mon. Not. Roy. Astron. Soc.\/} {\bf
  364} 635 [\eprint{astro-ph/0508105}]

\bibitem{Stergioulas:1994ea}
Stergioulas N and Friedman J 1995 {\em Astrophys. J.\/} {\bf 444} 306
  [\eprint{astro-ph/9411032}]

\bibitem{Paschalidis:2016vmz}
Paschalidis V and Stergioulas N 2016  [\eprint{1612.03050}]

\bibitem{Martinez:2015mya}
Martinez J~G, Stovall K, Freire P~C~C, Deneva J~S, Jenet F~A, McLaughlin M~A,
  Bagchi M, Bates S~D and Ridolfi A 2015 {\em Astrophys. J.\/} {\bf 812} 143
  [\eprint{1509.08805}]

\bibitem{Silva:2016myw}
Silva H~O, Sotani H and Berti E 2016 {\em Mon. Not. Roy. Astron. Soc.\/} {\bf
  459} 4378--4388 [\eprint{1601.03407}]

\bibitem{Popov:2006ki}
Popov S~B, Blaschke D, Grigorian H and Prokhorov M~E 2007 {\em Astrophys. Space
  Sci.\/} {\bf 308} 381 [\eprint{astro-ph/0606308}]

\bibitem{Yagi:2014qua}
Yagi K, Stein L~C, Pappas G, Yunes N and Apostolatos T~A 2014 {\em Phys.
  Rev.\/} {\bf D90} 063010 [\eprint{1406.7587}]

\bibitem{Stein:2013ofa}
Stein L~C, Yagi K and Yunes N 2014 {\em Astrophys. J.\/} {\bf 788} 15
  [\eprint{1312.4532}]

\bibitem{Baubock:2013gna}
Baubock M, Berti E, Psaltis D and Özel F 2013 {\em Astrophys. J.\/} {\bf 777}
  68 [\eprint{1306.0569}]

\bibitem{Gendreau2012}
Gendreau K~C, Arzoumanian Z and Okajima T 2012 {\em Proc.SPIE\/} {\bf 8443}
  8443 -- 8443 -- 8

\bibitem{Arzoumanian2014SPIE}
{Arzoumanian} Z, {Gendreau} K~C, {Baker} C~L, {Cazeau} T, {Hestnes} P,
  {Kellogg} J~W, {Kenyon} S~J, {Kozon} R~P, {Liu} K~C, {Manthripragada} S~S,
  {Markwardt} C~B, {Mitchell} A~L, {Mitchell} J~W, {Monroe} C~A, {Okajima} T,
  {Pollard} S~E, {Powers} D~F, {Savadkin} B~J, {Winternitz} L~B, {Chen} P~T,
  {Wright} M~R, {Foster} R, {Prigozhin} G, {Remillard} R and {Doty} J 2014 {The
  neutron star interior composition explorer (NICER): mission definition} {\em
  Space Telescopes and Instrumentation 2014: Ultraviolet to Gamma Ray\/} ({\em
  \procspie\/} vol 9144) p 914420

\bibitem{Zhang:2016ach}
Zhang S~N {\em et~al.\/} (eXTP) 2016 {\em Proc. SPIE Int. Soc. Opt. Eng.\/}
  {\bf 9905} 99051Q [\eprint{1607.08823}]

\bibitem{Doneva:2013rha}
Doneva D~D, Yazadjiev S~S, Stergioulas N and Kokkotas K~D 2013 {\em Astrophys.
  J.\/} {\bf 781} L6 [\eprint{1310.7436}]

\bibitem{Chakrabarti:2013tca}
Chakrabarti S, Delsate T, Gürlebeck N and Steinhoff J 2014 {\em Phys. Rev.
  Lett.\/} {\bf 112} 201102 [\eprint{1311.6509}]

\bibitem{Pappas:2013naa}
Pappas G and Apostolatos T~A 2014 {\em Phys. Rev. Lett.\/} {\bf 112} 121101
  [\eprint{1311.5508}]

\bibitem{Yagi:2014bxa}
Yagi K, Kyutoku K, Pappas G, Yunes N and Apostolatos T~A 2014 {\em Phys.
  Rev.\/} {\bf D89} 124013 [\eprint{1403.6243}]

\bibitem{Benhar:2005gi}
Benhar O, Ferrari V, Gualtieri L and Marassi S 2005 {\em Phys. Rev.\/} {\bf
  D72} 044028 [\eprint{gr-qc/0504068}]

\bibitem{Clifton:2011jh}
Clifton T, Ferreira P~G, Padilla A and Skordis C 2012 {\em Phys. Rept.\/} {\bf
  513} 1--189 [\eprint{1106.2476}]

\bibitem{Yunes:2013dva}
Yunes N and Siemens X 2013 {\em Living Rev. Rel.\/} {\bf 16} 9
  [\eprint{1304.3473}]

\bibitem{Berti:2015itd}
Berti E {\em et~al.\/} 2015 {\em Class. Quant. Grav.\/} {\bf 32} 243001
  [\eprint{1501.07274}]

\bibitem{Glampedakis:2015sua}
Glampedakis K, Pappas G, Silva H~O and Berti E 2015 {\em Phys. Rev.\/} {\bf
  D92} 024056 [\eprint{1504.02455}]

\bibitem{Glampedakis:2016pes}
Glampedakis K, Pappas G, Silva H~O and Berti E 2016 {\em Phys. Rev.\/} {\bf
  D94} 044030 [\eprint{1606.05106}]

\bibitem{Pani:2011xm}
Pani P, Berti E, Cardoso V and Read J 2011 {\em Phys. Rev.\/} {\bf D84} 104035
  [\eprint{1109.0928}]

\bibitem{Zaglauer:1992bp}
Zaglauer H~W 1992 {\em Astrophys. J.\/} {\bf 393} 685--696

\bibitem{Sotani:2017pfj}
Sotani H and Kokkotas K~D 2017 {\em Phys. Rev.\/} {\bf D95} 044032
  [\eprint{1702.00874}]

\bibitem{Hunter:2007}
Hunter J~D 2007 {\em Computing In Science \& Engineering\/} {\bf 9} 90--95

\end{thebibliography}

\providecommand{\newblock}{}

\end{document}